\documentclass[aps,prl,reprint,twocolum,floatfix,longbibliography]{revtex4-1}  
\usepackage{graphicx,graphics,color} 
\usepackage{amsmath,amssymb,amsfonts} 
\usepackage{bm,times,xspace} 
\usepackage[colorlinks,citecolor=blue,linkcolor=red]{hyperref} 
\begin{document}

\title{Observation of Geometric Heat Pump Effect in Periodic Driven Thermal Diffusion}

\author{Zi Wang}
\thanks{These two authors contributed equally to this work.}
\author{Jiangzhi Chen}
\thanks{These two authors contributed equally to this work.}
\author{Zhe Liu}
\author{Jie Ren}
\email{Corresponding Email: Xonics@tongji.edu.cn}
\affiliation{Center for Phononics and Thermal Energy Science, China-EU Joint Lab on Nanophononics, Shanghai Key Laboratory of Special Artificial Microstructure Materials and Technology, School of Physics Science and Engineering, Tongji University, Shanghai 200092, China
}%

\date{\today}

\begin{abstract}
The concept of geometry works as an overarching framework underlying a wide range of transport phenomena. Particularly, the geometric phase effect in classical and quantum heat pump has been attracting much attention in microscopic systems. 
Here, we formulate theoretically the geometric heat pump effect in macroscopic driven diffusive systems. Upon modulation protocols, the nontrivial geometric curvature in the parameter space universally induces an additional pumped heat, beyond the constraint of hot-to-cold flowing.
Furthermore, we set up a minimum experiment and indeed observe a non-vanishing directional heat flow across the driven system, despite keeping zero thermal bias between two time-dependent thermal reservoirs at every instant. We verify that in analogy to the geometric phase effect, the geometric pumped heat during each driving cycle is independent of driving periods in the adiabatic limit and coincides with theoretical predictions, thus validating its geometric origin. 
These results about geometric heat pump effect could have potential implications for designing and implementing nonreciprocal and topological thermal meta-devices under spatiotemporal modulations.
\end{abstract}

\maketitle

\emph{Introduction.}--
Recently the phononic thermal transport~\cite{dhar2008heat,li2012colloquium} including the search for functional thermal devices, such as thermal diodes~\cite{li2004thermal,chang2006solid}, thermal transistors~\cite{ben2014near,joulain2016quantum}, etc., raises a surge of interest, which is not only fundamentally central within the topics in nonequilibrium statistical physics~\cite{saito2007fluctuation,saryal2019thermodynamic,talkner2020colloquium}, but also pragmatically promises the next generation thermal computation and thermal energy control. Furthermore, a series of spatially constructed thermal metamaterials and meta-devices~\cite{yang2020controlling,li2021transforming} have also been proposed to delicately regulate the heat flow direction and magnitude.

Nevertheless, in static macroscopic configurations, heat always flows globally from hot sources to cold drains therein, as dictated by the second law of thermodynamics. To circumvent this constraint, people resort to temporal drivings, which have already been applied to generate non-trivial Floquet states both in closed~\cite{kitagawa2010topological,ma2018experimental} and open~\cite{engelhardt2019discontinuities,engelhardt2021dynamical} quantum systems. Meanwhile, temporal drivings have been introduced to thermal manipulations, producing novel theoretical effects like the directional heat flow in unbiased thermal transport~\cite{li2008ratcheting,ren2010emergence}, periodically driven heat engines~\cite{marathe2007two}, nonreciprocal thermal metamaterials~\cite{torrent2018nonreciprocal}, dynamic photonic refrigeration~\cite{buddhiraju2020photonic}, radiative heat shuttling~\cite{latella2018radiative}, adiabatic thermal radiation pump~\cite{li2019adiabatic}, etc., as well as experimental 
observations, such as high-performance solid-state electrocaloric cooling~\cite{wang2020high},  thermal non-Hermitian~\cite{li2019anti} and topological~\cite{xu2021configurable} dynamics, etc. These works explicitly demonstrate the power and versatility of temporal driving methods in thermal manipulation.

Considering the convenience brought by driving, it is demanding to grasp underlying universal concepts. The geometry emerges as one of the most insightful ideas. The geometric phase is originally proposed in closed quantum systems~\cite{berry1984quantal,thouless1983quantization}, later generalized to the scattering process in open quantum systems~\cite{brouwer1998scattering} and also to the full counting statistics~\cite{sinitsyn2007universal} in stochastic pump systems~\cite{astumian2002brownian,rahav2008directed} as an additional term obtained after a periodical modulation. Concerning the heat transport process, a similar geometric effect is unveiled in an anharmonic quantum junction~\cite{ren2010berry}, which induces an additional pumped heat from cold to hot after one periodic driving, so called geometric heat pump. Subsequently, the geometric heat pump effect raises a plethora of research on its manifestation in nano-sized open quantum~\cite{chen2013dynamic,wang2017unifying,nie2020berry} and classical coupled oscillators~\cite{ren2012geometric}. Also, recently the connections with the entropy production~\cite{sagawa2011geometrical}, heat engines~\cite{brandner2020thermodynamic,bhandari2020geometric,hino2021geometrical} and nonadiabatic control methods~\cite{funo2020shortcuts,takahashi2020nonadiabatic} are established. These works emphasize the central status of the geometric heat pump~\cite{wang2022geometric}.

Despite its massive theoretical attention and broad implications, the geometric heat pump effect has not been realized experimentally yet. Also, the geometric heat pump effect seems restricted to quantum nano-scale and microscopic stochastic systems, which brings the challenge obstructing the experimental verification. Therefore, a question naturally arises that whether the geometric heat pump effect is present in macroscopic diffusive systems. If so, how can the corresponding effect be experimentally observed and utilized to construct practical thermal devices? 

\begin{figure}[htbp]
 \centering
 \includegraphics[width=\linewidth]{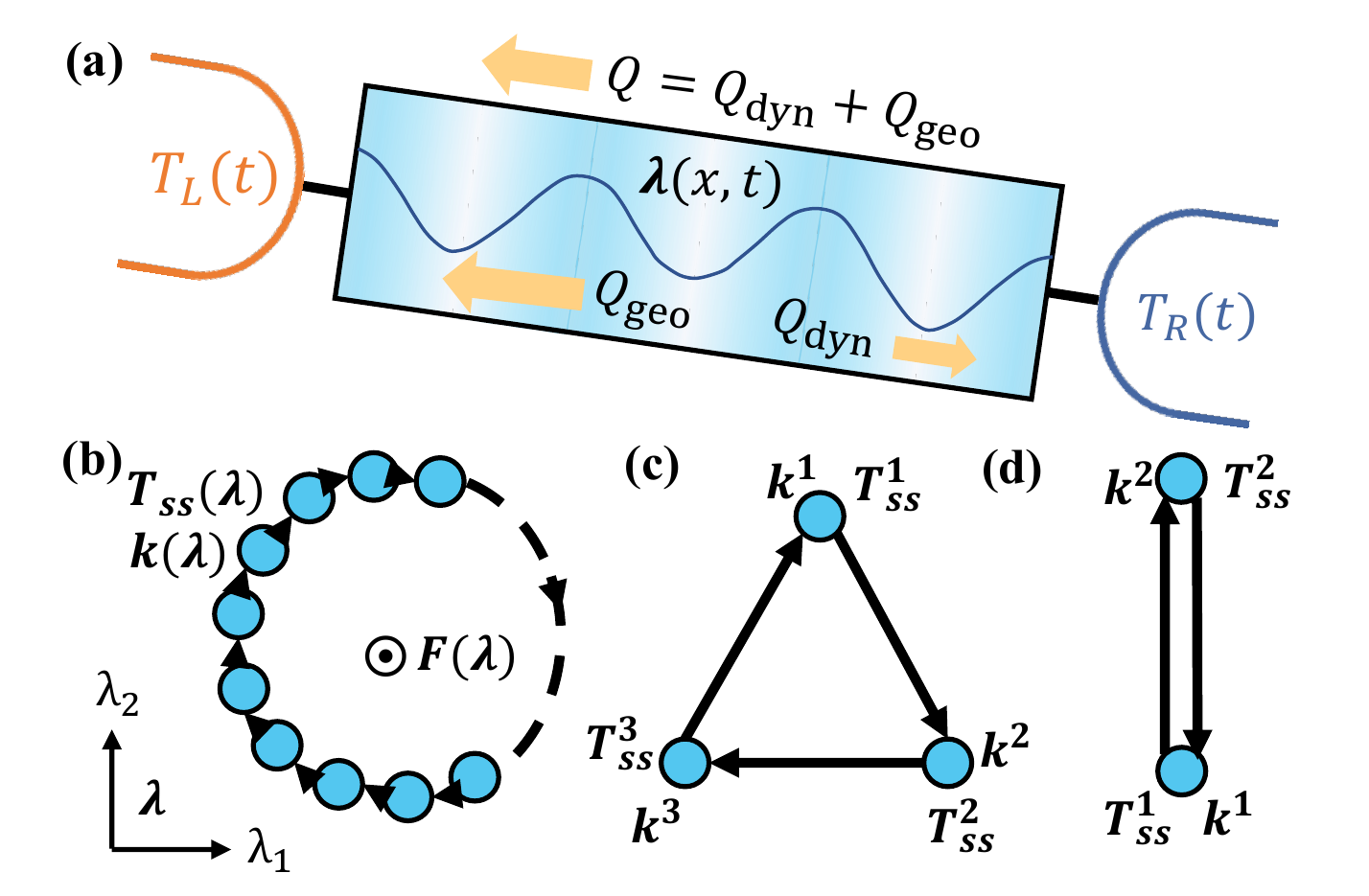}
 \caption{\textbf{A diagram of the geometry in driven diffusive systems from trotterized continuous protocols to discrete protocol.} (a) The heat transport as a competition between $Q_{\mathrm{geo}}$ and $Q_{\mathrm{dyn}}$. The dynamic heat $Q_{\mathrm{dyn}}$ is always parallel to the temperature bias $T_L(t)-T_R(t)$, but $Q$ can be against the bias, if the geometric heat $Q_{\mathrm{geo}}$ is dominant. (b) The trotterized continuous driving path. Filled dots $\boldsymbol{\lambda} \equiv (\lambda_1,\lambda_2)$ denote points in the parameter space. $Q_{\mathrm{geo}}$ is induced by the intrinsic nontrivial curvature $\mathbf{F}(\boldsymbol{\lambda})$. Both instantaneous steady state temperature $\mathbf{T}_{ss}$ and dual $\mathbf{k}$ are defined locally. (c) The multiple-point protocol defined by trotterizing the continuous protocol. The inner product between $\mathbf{k}^n$ and $\Delta T_{ss}^n \equiv T_{ss}^{n}-T_{ss}^{n-1}$ quantifies geometric contribution during relaxation from steady state at point $n-1$ to that at $n$, wherein the systems' parameters are fixed at $\boldsymbol{\lambda}^n$. (d) The two-point switching model. It corresponds to our experimental setup. }
 \label{fig1}
 \end{figure}

In this Letter, we draw a positive conclusion on above questions. We excavate the geometric heat contribution in addition to the conventional dynamic heat flow in general driven diffusive systems, incorporating the situations of both continuous and discrete driving protocols, as shown in Fig.~\ref{fig1}. We further elaborate a periodic driven experimental setup with reservoirs' temperature and thermal conductance modulated to demonstrate the geometric heat pump effect in the macroscopic thermal diffusion. The experimental observation verifies that a non-zero geometric heat flow is generated even under instantaneous zero bias at every instant. Meanwhile, the integrated heat contribution over each single driving period in the adiabatic limit (long driving period) is independent of the driving period, clearly validates our theoretical geometric formulation.

\emph{Geometric Heat Pump Effect.}--We begin by recapitulating the well-known classical thermal diffusion, governed by the Fourier's law~\cite{callen1998thermodynamics} $\mathbf{J}=-\kappa \nabla T(x,t)$ and the continuity equation $\frac{\partial}{\partial t} E=-\nabla \cdot \mathbf{J}$. The Fourier's law states that the heat current density $\mathbf{J}$ is proportional to the conductivity $\kappa$ and inverse temperature gradient, while the continuity condition simply accounts for the energy conservation.  

Spatially discretizing the central system with temperature vector $\mathbf{T}_c$ and take into account the boundary temperature vector $\mathbf{T}_b$ fixed by the connected reservoirs, the heat flow during the linear conduction process is given by the discrete Fourier's law: $\mathbf{J}=\mathcal{K}_c \mathbf{T}_c +\mathcal{K}_b \mathbf{T}_b$, where the conduction matrix $\mathcal{K}_c$ ($\mathcal{K}_b$) depicts the conduction effect induced within system itself (by coupling to thermal reservoirs). The heat flow $\mathbf{J}$ is a column vector containing currents between different neighbouring space points in the whole system. Considering the continuity relation governed by the energy conservation, the heat flow incurs the temperature evolution as $\frac{\partial}{\partial t}\mathbf{T}_c=\mathcal{L}\mathbf{J}$. Here, $\mathcal{L}\equiv \mathcal{C}^{-1}\mathcal{D}$. The diagonal matrix $\mathcal{C}$ contains the heat capacity as its elements and $\mathcal{D}$ is the divergence matrix (the discrete version of divergence operator $\nabla \cdot$). Their concrete form is exemplified in the section 1 of supplement~\cite{supp}.

Therefore, by combining the above two equations, the evolution of the system's temperature distribution is described by
\begin{equation}
\label{temp_evolution}
    \frac{\partial}{\partial t}\mathbf{T}_c=\mathcal{M}_c \mathbf{T}_c +\mathcal{M}_b \mathbf{T}_b, 
\end{equation}
where $\mathcal{M}_c \equiv \mathcal{L} \mathcal{K}_c$ and $\mathcal{M}_b \equiv \mathcal{L} \mathcal{K}_b$. The invertible matrix $\mathcal{M}_c$ describes the cooling process of the central system if the reservoirs are set to the zero temperature. 
The above abstract notions are exemplified by a concrete one dimensional model (See section 1 in supplement~\cite{supp}) to clarify the physical implication. 

In the adiabatic limit, the driving is slow enough for the system to relax to its instantaneous steady state at every instant. According to the adiabatic perturbation theory~\cite{kolodrubetz2017geometry}, $\partial \mathbf{T}_c/\partial t$ can be considered as a perturbation term. To the zeroth order, the steady state of $\mathbf{T}_c$ is obviously $\mathbf{T}_{ss}=-(\mathcal{M}_c)^{-1}\mathcal{M}_b \mathbf{T}_b$. This is the exact result of $\mathbf{T}_c(t)$ if no driving is applied. 
To the first order of driving speed, adiabatic perturbation (as shown in the section 2 of the supplement~\cite{supp}) yields the separation of heat flow into the dynamic and geometric components
\begin{equation}
\label{dynamical}
    J_{\mathrm{dyn}}=\mathbf{1} \cdot \mathcal{K}_c \mathbf{T}_{ss}+\mathbf{1} \cdot \mathcal{K}_b \mathbf{T}_b, 
\end{equation}
\begin{equation}
\label{geometric}
    J_{\mathrm{geo}}=\mathbf{k} \cdot \frac{\partial \mathbf{T}_{ss}}{\partial \lambda_i}\dot{\lambda}_i, 
\end{equation}
where we adopt the convention of summing repeated indices. The vector $\mathbf{k} \equiv \mathbf{1} \cdot \mathcal{K}_c (\mathcal{M}_c)^{-1}$ is a row vector mapping the variation speed of $\mathbf{T}_{ss}$ into heat current components. It is in the dual space of $\mathbf{T}_{ss}$ and is reminiscent of the quantum 
``bra". $\mathbf{1}$ is a row vector whose components are zero except $\mathbf{1}_i=1$, with $i$ denoting our interested current component. $\boldsymbol{\lambda}$ is the vector of driven parameters. The dynamic part is simply a steady state current in static situation. The geometric contribution $J_{\mathrm{geo}}$ has no analogy in static systems. 

Suppose the parameters of the system are driven cyclically, the accumulative geometric heat flow during one single period is
\begin{equation}
    Q_{\mathrm{geo}}\equiv \int_0^{\tau_p}J_{\mathrm{geo}}(t)dt=\oint_{\partial \Omega} d \boldsymbol{\lambda} \cdot \mathbf{A}. 
\end{equation}
Here, the geometric connection independent of the driving speed is $\mathbf{A} \equiv \mathbf{k} \cdot \nabla_{\boldsymbol{\lambda}} \mathbf{T}_{ss}$, with $\partial \Omega$ being the closed protocol path. It is reminiscent of the original Berry connection~\cite{berry1984quantal} and similar counterparts in microscopic stochastic systems~\cite{ren2010berry}. The protocol can be parametrized as $\{ \boldsymbol{\lambda}(t) \}_{0 \leq t < \tau_p}$, where $\tau_p$ is the driving period. 

In the situation where two parameters are modulated, i.e., $\boldsymbol{\lambda}=(\lambda_1,\lambda_2)^\mathrm{T}$, $Q_{\mathrm{geo}}$ is also formulated as, with the aid of the Stokes formula, 
\begin{equation}
\label{continuous_geo}
\begin{split}
&  Q_{\mathrm{geo}}=\int_{\Omega} F(\boldsymbol{\lambda}) d\Omega, \\
& F(\boldsymbol{\lambda}) \equiv \frac{\partial \mathbf{k}}{\partial \lambda_1} \cdot  \frac{\partial \mathbf{T}_{ss}}{\partial \lambda_2}-\frac{\partial \mathbf{k}}{\partial \lambda_2} \cdot  \frac{\partial \mathbf{T}_{ss}}{\partial \lambda_1}. 
\end{split}
\end{equation}
The curvature $F(\boldsymbol{\lambda})$ is the geometric curvature and the integration is over the area $\Omega$ encircled by the closed path $\partial \Omega$. $d \Omega$ is the corresponding surface element. The intrinsic nontrivial curvature $F(\boldsymbol{\lambda})$ induces the accumulated pumped heat in each period, independent of $\tau_p$. 

\emph{Switching Driving Protocol.}--A continuous driving protocol can be trotterized into a discrete one. Here we discuss the crossover from continuous drivings to protocols where system parameters are just switched back and forth between several states.  Accordingly, the dynamic and geometric components of the accumulated pumped heat as integrals are replaced by corresponding summations 
\begin{equation}
    Q_{\mathrm{dyn}}=\sum_{n=1}^{N}(\mathbf{1} \cdot \mathcal{K}_{c}^n \mathbf{T}_{ss}^n+\mathbf{1} \cdot \mathcal{K}_{b}^n \mathbf{T}_{b}^n)\frac{\tau_p}{N}, 
\end{equation}
\begin{equation}
\label{discrete_geo}
    Q_{\mathrm{geo}}=\sum_{n=1}^{N} \mathbf{k}^n \cdot \Delta \mathbf{T}_{ss}^n. 
\end{equation}
$Q_{\mathrm{geo}}$ has a nature comparable with the original Pancharatnam geometric phase~\cite{pancharatnam1956generalized}, but here in a non-Hermitian sense. We trotterize the protocol into $N$ pieces, with system parameters being constant within each $\tau_p/N$ duration. $\mathcal{K}_{c}^n$, $\mathcal{K}_{b}^n$ and $\mathbf{k}^n$ correspond to the system parameters in the $n$-th protocol piece. After each quench of system parameters, the system relaxes and dissipates. Considering the adiabaticity of driving, the system temperature relaxes to its steady state at the end of each driving piece. $\mathbf{T}_{ss}^n$ is the system temperature after finishing the $n$-th piece and right before the next quench. Also, $\mathbf{T}_{ss}^n$ is calculated to be $\mathbf{T}_{ss}^n \equiv -(\mathcal{M}_{c}^n)^{-1}\mathcal{M}_{b}^n \mathbf{T}_{b}^n$, identical to the adiabatic steady state in the continuous driving situation. In Eq.~(\ref{discrete_geo}), the difference between two adjacent steady states is defined as $\Delta \mathbf{T}_{ss}^n \equiv \mathbf{T}_{ss}^n-\mathbf{T}_{ss}^{n-1}$. For the details of a direct derivation, see section 2 in supplement~\cite{supp}. 

As a limit case, also in alignment with the experiment carried out and shown below, the geometrically pumped heat is simplified to be
\begin{equation}
\label{eq12}
    Q_{\mathrm{geo}} =(\mathbf{k}^1-\mathbf{k}^2) \cdot (\mathbf{T}_{ss}^1-\mathbf{T}_{ss}^2), 
\end{equation}
if the system parameters are cyclically switched between only two states. This is apparently independent of $\tau_p$, justifying its geometric origin. The above result encapsulates a non-zero directional heat flow arising solely from the driving even when no instantaneous non-zero thermal bias is introduced. This phenomenon is observed in our experiment. A unified geometric perspective of continuous and discrete protocols is schematically shown in Fig.~\ref{fig1}. 

\begin{figure}[htbp]
\centering
\includegraphics[width=\linewidth]{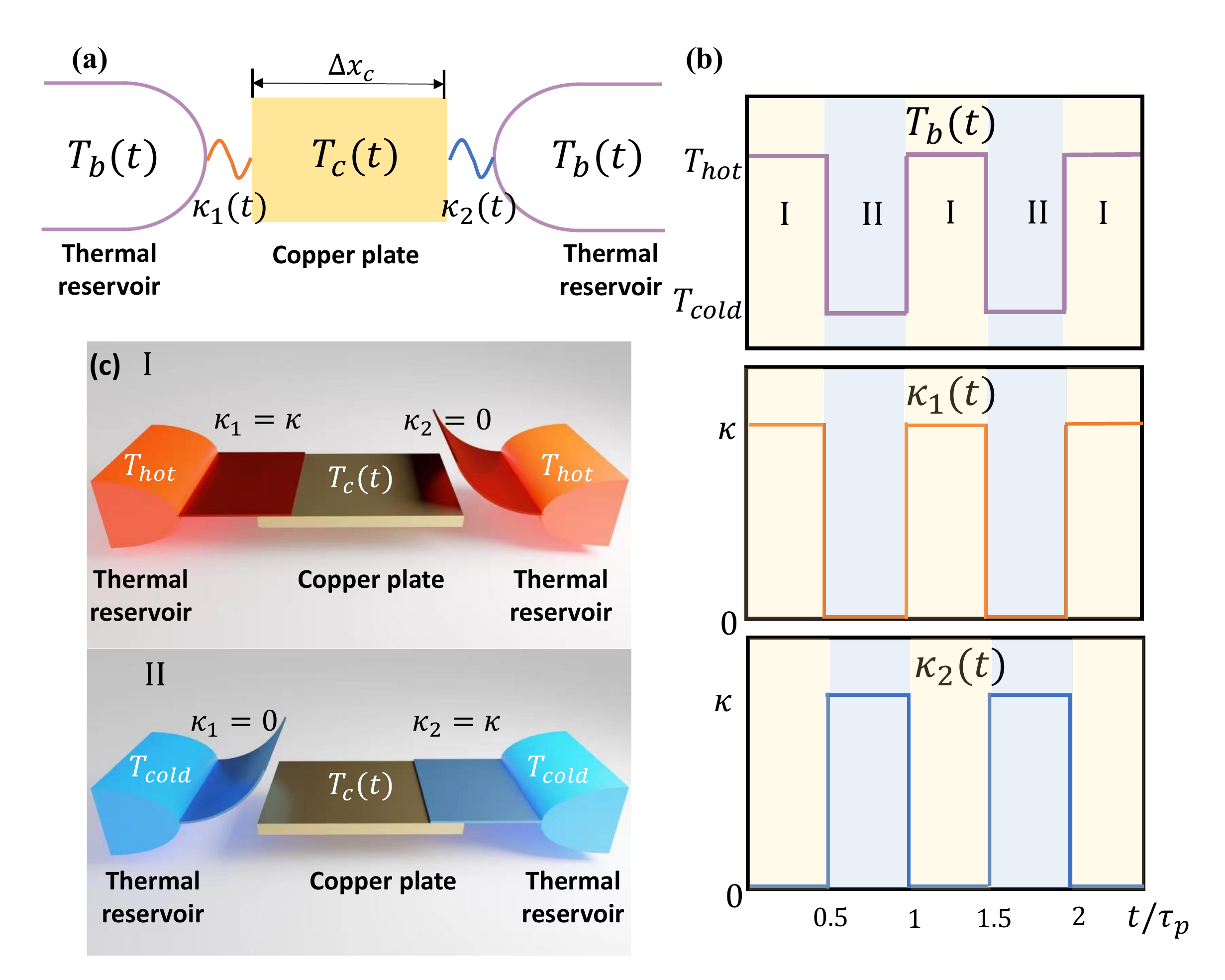}
\caption{\textbf{The schematic diagram of the experimental setup and the driving protocol.} (a) The central system (the copper plate) is coupled to two thermal reservoirs with identical temperature. The temperature of the copper plate is $T_{c}$ and that of reservoirs is $T_{b}$. The two thermal conductances between the copper plate and reservoirs are $\kappa_{n}$ ($n=1,2$) with a maximum $\kappa$. $T_{c}$, $T_{b}$ and $\kappa_{n}$ change over time. $\Delta x_{c}$ is the width of the copper plate. (b) The temporal modulation of the reservoirs' temperature $T_b$ and conductance $\kappa_n$ with $\tau_{p}$ being the driving period. The system parameters are switched instantaneously between state \uppercase\expandafter{\romannumeral1} and \uppercase\expandafter{\romannumeral2}. (c) The corresponding system configurations in two parameter states (\uppercase\expandafter{\romannumeral1} and \uppercase\expandafter{\romannumeral2}). The conductance modulation is controlled by the SMA.}
\label{fig2}
\end{figure}

\emph{Experimental Demonstration.}---To observe the geometric heat pump, we start with a typical structure as described in Fig.~\ref{fig2}(a). We connect both ends of the copper plate to the thermal reservoirs with identical temperatures $T_{b}$. The thermal conductance between the copper plate and two reservoirs, i.e. $\kappa_{1}$ and $\kappa_{2}$, are out of phase. In Fig.~\ref{fig2}(b), we show the 
periodic protocols of $T_{b}$, $\kappa_{1}$ and $\kappa_{2}$. We demonstrate the two possible system configurations in Fig.~\ref{fig2}(c). In configuration \uppercase\expandafter{\romannumeral1}, the thermal reservoirs are at high temperature, and the copper plate is only connected to the left thermal reservoirs. The thermal conductance on the left equals its maximum value $\kappa$. From \uppercase\expandafter{\romannumeral1} to \uppercase\expandafter{\romannumeral2}, we instantaneously switch the temperature of reservoirs from $T_{\mathrm{hot}}$ to $T_{\mathrm{cold}}$. Meanwhile, the central system is connected to the right reservoir and disconnected from the left one. 

To modulate $\kappa_1$ and $\kappa_2$ in the experiment, we construct the system-reservoir couplings with two types of reversible shape memory alloys (SMA). The geometric configurations of SMA change rapidly as its temperature varies, which is also applied in macroscopic thermal diodes~\cite{PhysRevLett.115.195503}. When the thermal reservoirs are at $T_{\mathrm{hot}}$, the SMA-1 is flat and the SMA-2 is warped, and when the reservoirs are switched to $T_{\mathrm{cold}}$, the configurations of two SMAs are reversed. Therefore, the switching of reservoirs spontaneously induces the modulation of $\kappa_1$ and $\kappa_2$, through the mechanical response of SMAs. 

\begin{figure}[htbp]
\centering
\includegraphics[width=\linewidth]{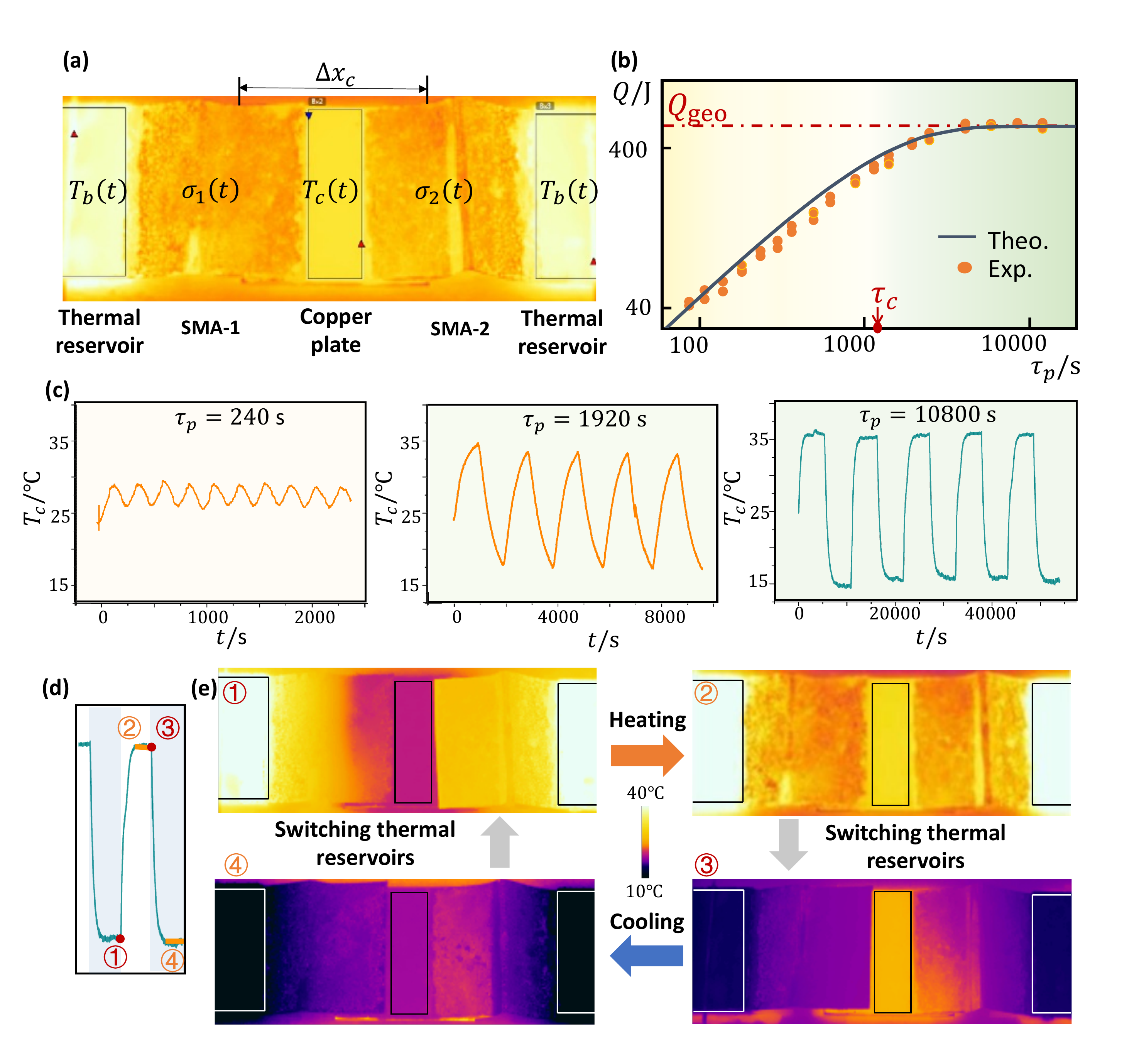}
\caption{\textbf{The observation of the geometric heat pump in the thermal diffusion system.} (a) A snapshot of the experimental device generated by an infrared thermal imager, corresponding to Fig.~\ref{fig2}. (b) The heat transferred during a single driving period versus different $\tau_{p}$. The characteristic time $\tau_{c}\equiv$ $2C/\sigma$=1070s signifies the crossover between two regimes--fast driving and slow driving, marked by the yellow and green background color, respectively. The geometric heat pump effect is observed as the plateau $Q_\mathrm{geo}$=546J when the driving is slow. (c) The evolution of $T_{c}$ in different $\tau_{p}$ situations, corresponding to the fast, intermediate, and fast driving cases in (b). The maximum and minimum of $T_{c}$ get closer to that of the thermal reservoirs as $\tau_{p}$ gets longer. (d) A typical period in the evolution of $T_c$  ($\tau_{p}$=10800s), which contains four phases. The background colors distinguish the different parameter states. (e) Snapshots of the experiment device in the four phases in the adiabatic limit. The four phases correspond to \textcircled{\small{1}}, \textcircled{\small{2}}, \textcircled{\small{3}} and \textcircled{\small{4}} in (d), respectively. }
\label{fig3}
\end{figure}

Figure~\ref{fig3}(a) is an infrared imaging snapshot of Fig.~\ref{fig2}(a). The copper plate (with a width of $\Delta x_c$=50mm) is coupled to two thermal reservoirs with heat capacity much greater than that of the plate. The SMAs are much thinner than the copper plate (see section 3 in supplement~\cite{supp} for more details of the setup), rendering its capacity negligible. 
The thermal conductance $\kappa_{n}$ in our experiment is not necessarily used for analysis, due to the presence of dissipation into the surrounding atmosphere. Thus, we calibrate the effective conductances $\sigma_{n}$ instead of $\kappa_{n}$ in the experiment. We couple the copper plate to one side of thermal reservoir at a time and record the time dependence of $T_c$. $\sigma$=0.049 W/K is obtained from the mean relaxation time (see more discussions in section 4 of supplement~\cite{supp}). We also calibrate the maximum and minimum temperature of the effective thermal reservoir $T_e$ as $T_{\mathrm{hot}}=35.2^{\circ}$C and $T_{\mathrm{cold}}=14.8^{\circ}$C (section 4 of supplement~\cite{supp}). 

According to the theory, we analytically calculate the heat transferred during one single period as
\begin{equation}
    Q=C\Delta T_{e}(1-e^{-\frac{\sigma \tau_p}{2C}}),
\label{eq13}
\end{equation}
where $C$ is the heat capacity of the central system, $\Delta T_{e}\equiv T_{\mathrm{hot}}-T_{\mathrm{cold}}$ (more details in section 5 of supplement~\cite{supp}). In our experimental setup, $C$=26.208 J/K and $\Delta T_{e}$=20.4$^{\circ}$C. In spite of the driving frequency, the dynamical component of the transferred heat during one single period is $Q_\mathrm{dyn}\equiv0$ due to the strict instantaneous zero bias of the thermal reservoirs. The pumped heat is singly contributed to $Q_{\mathrm{geo}}$. 

In the adiabatic limit ($\tau_{p}\to \infty$), Eq.~(\ref{eq13}) reduces to
\begin{equation}
    Q = C\Delta T_{e}. 
\label{eq14}
\end{equation}
To form a connection with our general theory, we note that steady states $\mathbf{T}_{ss}^n := T_b^n$ ($n=$\uppercase\expandafter{\romannumeral1}, \uppercase\expandafter{\romannumeral2}). Also, defining the positive direction of the current as left to right, $\mathcal{M}_c^n:=-(\sigma_1^n+\sigma_2^n)/C$ and $\mathcal{K}_c^n:=(-\sigma_1^n,\sigma_2^n)^\mathrm{T}$. Without losing any generality, we select $\mathbf{1}:=(1,0)^\mathrm{T}$. Thus, $\mathbf{k}^1-\mathbf{k}^2=C$ and  Eq.~(\ref{eq12}) reduces to $Q_\mathrm{geo}=C\Delta T_{e}$, in agreement with Eq.~(\ref{eq14}), showing $Q$ in our experiment to be purely geometric. The details of this compare is given in the section 5 of supplement~\cite{supp}. Also, the measured result of Eq.~(\ref{eq14}) is shown in Fig.~\ref{fig3}(b). 

Figure~\ref{fig3}(b)(c) shows results of our observation and measurement of the geometric heat pump effect. We record the temperature evolution of the copper plate $T_{c}(t)$ under different driving period $\tau_{p}$. After several periods, the system enters a periodic state. Then we measure the heat transferred during one single period under different driving periods $\tau_{p}$, which is displayed in Fig.~\ref{fig3}(b). The black line shows the theoretical value from Eq.~(\ref{eq13}) and the dots show the experimental results. In the fast driving region, $Q$ and $\tau_{p}$ are approximately linearly correlated, consistent with $Q\approx \frac{\sigma \Delta T_e}{2}\tau_p$. In the adiabatic limit ($\tau_p \gg \tau_c\equiv2C/\sigma$), $Q$ remains unchanged as $\tau_{p}$ varies, and its experimental plateau value $Q_\mathrm{geo}=545$J agrees with the theoretical value $Q_\mathrm{geo}=535$J predicted by Eq.~(\ref{eq14}). This proves the existence of general geometric heat pump effects in the classical diffusive transport. 

To further analyze the pump process, Fig.~\ref{fig3}(c) shows the variations of the copper plate's temperature $T_{c}$ in different driving periods $\tau_{p}$, representing the two cases--the fast driving (yellow background) and slow driving (green background) regimes. In the case of fast driving ($\tau_{p}$=240 s), $T_{c}$ reciprocates in a small interval due to the system's being impossible to respond instantaneously. As $\tau_{p}$ increases to 1920s, the difference of $T_{c}$ in a period gets larger. Within the adiabatic regime ($\tau_{p}$=10800 s), the maximum (minimum) value of $T_c(t)$ is equal to $T_{\mathrm{hot}}$ ($T_\mathrm{cold}$). Compared with the instantaneous change of the thermal reservoirs, the delay of $T_{c}$ owes to the finite relaxation time of thermal diffusion. 

In the adiabatic limit ($\tau_{p}$=10800 s), the thermodynamic process is formed of four phases in one period, which are marked as \textcircled{\small{1}}, \textcircled{\small{2}}, \textcircled{\small{3}}, \textcircled{\small{4}} in Fig.~\ref{fig3}(d). We display the details of the four phases with infrared imaging snapshots in Fig.~\ref{fig3}(e). From phase \textcircled{\small{1}} to \textcircled{\small{2}}, the central system is coupled to the left hot thermal reservoir. Heat flows from left reservoir to the central system. After switching the two hot thermal reservoirs to cold ones, the central system only connects with the right reservoir. Heat flows from the central system to the right thermal reservoir (phase \textcircled{\small{3}} to \textcircled{\small{4}}). Thus, there is a one-way heat flow in the device. This thermodynamic cycle analysis provides an intuitive picture of the geometric heat pump effect. Since the central system is always isolated with one of the reservoirs, the pumping can be unambiguously traced to the difference of central system's two steady states, which equates $\Delta T_e$. The corresponding absorbing from the left (draining into the right) incurs the central system's energy change~[Eq. (\ref{eq14})] from phase \textcircled{\small{1}} to \textcircled{\small{2}} (\textcircled{\small{3}} to \textcircled{\small{4}}). This energy change is evidently the pumped heat during the whole cycle and independent of the adiabatic driving period. This attributes the physical picture of the geometrically pumped heat, i.e.~Eq. (\ref{continuous_geo})(\ref{discrete_geo})(\ref{eq12}), to the changing of system states, which originates to the intrinsic geometric properties of the parameter space.

\emph{Conclusion.}---To summarize, we have unveiled the intrinsic geometry underlying the driven diffusive thermal conduction process through both theoretical formulation and experimental observation. The experimental result in the adiabatic limit is well described by our derived general theoretical results. Our work, by unambiguous observation of the geometric heat flow, paves the way for future generalization to more complex systems. This would enable a versatile and reliable resource in generating directional heat flow and harnessing thermal energy by devising a dynamical toolkit. Constructing geometric effect induced non-reciprocal and topological  phenomena~\cite{torrent2018nonreciprocal,fernandez2021extreme} in driven diffusive processes constitutes a promising task. 

\begin{acknowledgments}
\textit{Acknowledgments}. We acknowledge the support from the National Natural Science Foundation of China (Nos. 11935010 and 11775159), and the Opening Project of Shanghai Key Laboratory of Special Artificial Microstructure Materials and Technology.
\end{acknowledgments}

\bibliography{ref.bib}

\end{document}